%% file: main.tex
\documentclass[a4paper,fleqn]{cas-sc}
\usepackage{aas_macros,xspace}
\usepackage[greek,english]{babel}
\usepackage{newunicodechar}
\usepackage[version=3]{mhchem}

\usepackage[square, numbers]{natbib}

\def\tsc#1{\csdef{#1}{\textsc{\lowercase{#1}}\xspace}}
\tsc{WGM}
\tsc{QE}

\newcommand{\mycounter}[2]{
\newcounter{#1}
\setcounter{#1}{#2}
}
\DeclareRobustCommand{\okina}{%
  \raisebox{\dimexpr\fontcharht\font`A-\height}{%
    \scalebox{0.8}{`}%
  }%
}
\newunicodechar{ʻ}{\okina}
\newcommand{\oumuamua}{ʻOumuama\xspace}

\mycounter{Npapers}{0}
\mycounter{searches}{16}
\addtocounter{Npapers}{\thesearches}
\mycounter{methods}{11}
\addtocounter{Npapers}{\themethods}
\mycounter{targets}{4}
\addtocounter{Npapers}{\thetargets}
\mycounter{development}{17}
\addtocounter{Npapers}{\thedevelopment}
\mycounter{theory}{26}
\addtocounter{Npapers}{\thetheory}
\mycounter{social}{6}
\addtocounter{Npapers}{\thesocial}

\begin{document}
\let\WriteBookmarks\relax
\def\floatpagepagefraction{1}
\def\textpagefraction{.001}
\defcitealias{Tusay2022}{Tusay \& Huston et al.}

\shorttitle{SETI in 2022}    

\shortauthors{Wright et al.}  

\title[mode = title]{SETI in 2022}  



%



\ead{astrowright@gmail.com}

\ead[url]{https://sites.psu.edu/astrowright/}


\affiliation[1]{organization={Department of Astronomy \& Astrophysics, Penn State}, addressline={525 Davey Lab}, city={University Park}, postcode={16802}, state={PA}, country={USA}}
\affiliation[2]{organization={Penn State Extraterrestrial Intelligence Center, Penn State}, addressline={525 Davey Lab}, city={University Park}, postcode={16802}, state={PA}, country={USA}}
\affiliation[3]{organization={Center for Exoplanets and Habitable Worlds, Penn State}, addressline={525 Davey Lab}, city={University Park},  postcode={16802}, state={PA}, country={USA}}
\affiliation[4]{organization={Department of Astronomy, University of California}, addressline={501 Campbell Hall}, city={Berkeley}, postcode={94720}, state={CA}, country={USA}}
            
\author[1,2,3]{Jason T. Wright}[orcid=0000-0001-6160-5888]
\cormark[1]






\author[2,4]{Macy Huston}[orcid=0000-0003-4591-3201]

\author[1,2,3]{Aidan Groenendaal}[orcid=0009-0001-1303-8898]



\author[1,2,3]{Lennon Nichol}[orcid=0009-0003-0255-2327]

\author[1,2,3]{Nick Tusay}[orcid=0000-0001-9686-5890]



\begin{abstract}
In this third installment of SETI in 20xx, we very briefly and subjectively review developments in SETI in 2022. Our primary focus is \theNpapers\ papers and books published or made public in 2022, which we sort into six broad categories: results from actual searches, new search methods and instrumentation, target and frequency selection, the development of technosignatures, theory of ETIs, and social aspects of SETI.
\end{abstract}




\maketitle

\input{text.tex}

\section*{Acknowledgements}
We thank Manasvi Lingam for reviewing this manuscript and providing several additional references we had missed.

This research has made use of NASA's Astrophysics Data System Bibliographic Services. This research was supported by the Center for Exoplanets and Habitable Worlds and the Penn State Extraterrestrial Intelligence Center, which are supported by Penn State and the Eberly College of Science.

\bibliographystyle{cas-model2-names}


\input{output.bbl}
\bio{}
\endbio

\bio{}
\endbio

\end{document}

%% file: text.tex
\section{Introduction} \label{sec:intro}

This is the third installment of the SETI in 20xx a series, begun by \citet{SETIin2020}, and continuing in \citet{Huston2022}, which aims to survey the additions to the SETI literature each year. As with previous installments, we aim to be ``usefully subjective'' in our choices of what to include and discuss and do not claim to be exhaustive. 

Our methodology for identifying papers is an update on the the methodology of \citet{Reyes2019} and \citet{LaFond2021}, which we use to produce the monthly mailers at \href{http://seti.news}{seti.news}, and the SETI bibliography at ADS. For a thorough list of all bibliographic entries tracked by ADS, one can use the \texttt{bibgroup:SETI} and \texttt{year:2022} search query elements there\footnote{That is, enter  \href{https://ui.adsabs.harvard.edu/search/q=bibgroup\%3ASETI\%20year\%3A2022&sort=date\%20desc\%2C\%20bibcode\%20desc&p_=0}{bibgroup:SETI year:2022} in the ADS search tool.}.

For consistency, we categorize the papers from 2022 into the same categories as in our previous installments, and structure this review in the same way.  Our categories are:

\begin{itemize}
    \item \textbf{Searches} (S): Actual searches for ETI, with focus on collecting and/or interpreting data and calculating upper limits.
    \item \textbf{Instrumentation and Methodology} (IM): Discussion of hardware and software for collecting and interpreting data.  
    \item \textbf{Target and Frequency Selection} (TF): The when, where, and how to look.  This includes discussions of Schelling points and other considerations for where in search space technosignatures might be found.
    \item \textbf{Development of Technosignatures} (DT): Proposals for new technosignatures, calculations of the observable consequences of technosignatures, or discussion of the form technosignatures might take, such as message composition. 
    \item \textbf{Theory of ETIs} (T): Discussions of the Fermi Paradox, Drake Equation, Galactic settlement and the prevalence of ETIs, the Great Filter, the statistics of one, and other purely theoretical considerations.
    \item \textbf{Social Aspects} (SA): Post-detection protocols, the ethics of METI, and similar focus on the human aspects of SETI.
\end{itemize}

Because of the interplay of instrumentation, methodology, theory, and observation, our categories are porous and many papers straddle topics or evade categorization entirely. For simplicity and consistency we nonetheless choose a single category for each paper. While we roughly group our papers in the order given above, in a few cases we group papers by topic instead (as with \oumuamua papers).  In these cases, we indicate which papers are in the ``wrong'' section using the parenthetical abbreviations above.

Additionally, we acknowledge that our system is imperfect and some relevant publications may be missed for a variety of reasons including timing of publication or access to a particular journal. We have reviewed our process and we include the articles missed in our previous installment in the final category ``Missed in 2021.''

\section{SETI in 2022}

This year saw a mild decrease in SETI publications from 2021.  The ADS SETI bibliography lists 112 entries in 2020, 155 in 2021, and 130 in 2022; the number of refeereed papers in those years held approximately steady at 63, 68, and 65 respectively.  This review with \theNpapers\ entries has fewer entries than the 2021 instance (98, not counting the 9 we add here), but more than the 2020 instance (75).  

2022 saw the return of in person conferences.

\begin{figure}
    \centering
    \includegraphics[width=\textwidth]{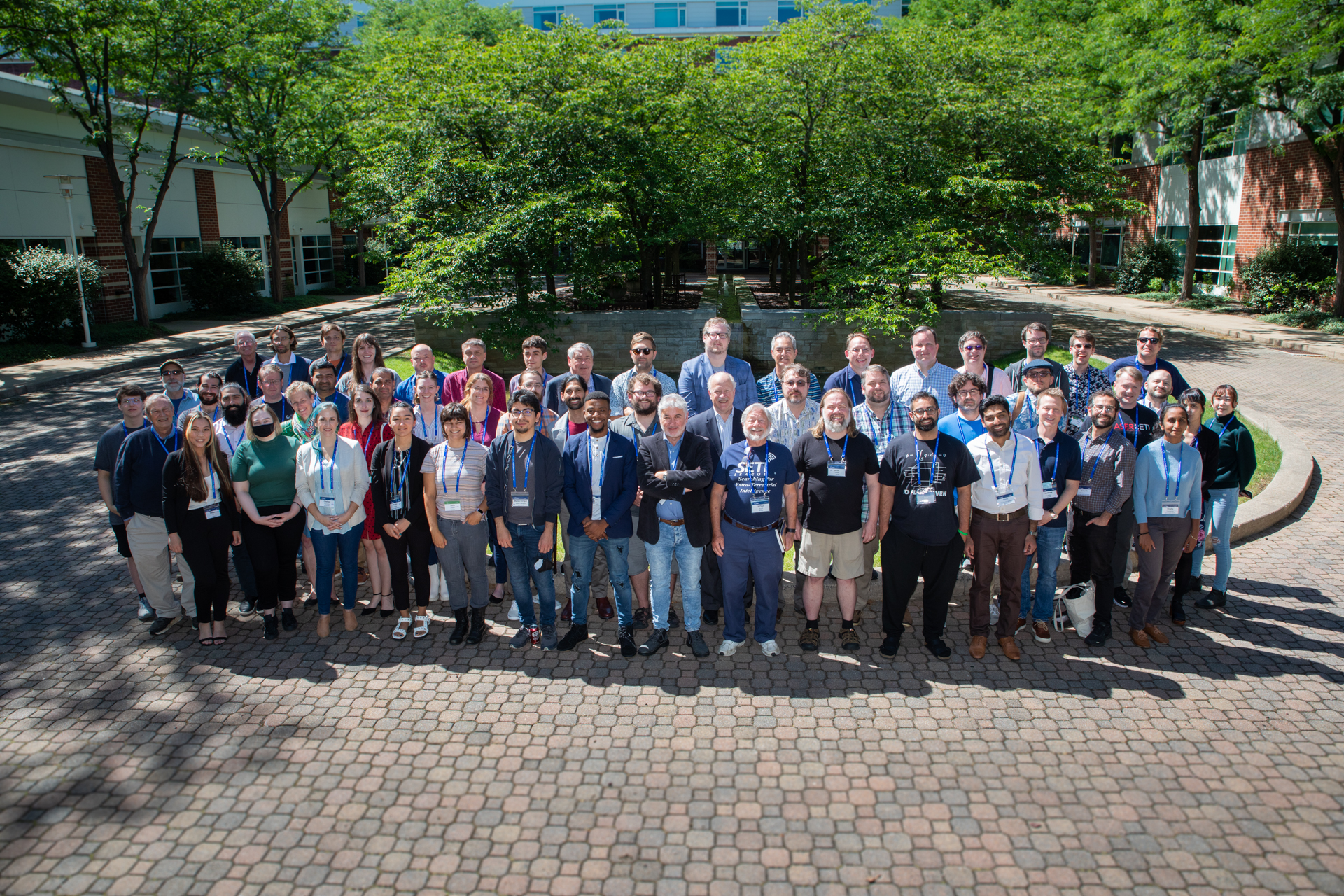}
    \caption{In-person attendees of the First Penn State SETI Symposium. Photo credit: Michael Fleck.}
    \label{fig:SETISymposium}
\end{figure}

The long-delayed Penn State SETI Symposium was held June 27--30 in State College Pennsylvania.  Over fifty people attended in person (Figure~\ref{fig:SETISymposium}) with another forty or so online.  The conference included plenary sessions and breakout sessions, and spanned the entirety of the field.  The talks and posters are available in the conference Zenodo community.\footnote{\href{https://zenodo.org/communities/pseti-symposium-2022/records?q=&l=list&p=1&s=10&sort=newest}{\url{https://zenodo.org/communities/pseti-symposium-2022/records?q=&l=list&p=1&s=10&sort=newest}}}

The 44th Scientific Assembly of COSPAR was held July 16--24 in Athens included a session on SETI with over 14 presenters.\footnote{The abstracts can be found on ADS via the search query \href{https://ui.adsabs.harvard.edu/search/q=bibstem:"cosp" bibgroup:SETI year:2022}{\texttt{bibstem:"cosp" bibgroup:SETI year:2022}}
}

The 2022 Drake Award, bestowed by the SETI Institute, went to Shelley Wright of UC San Diego for her pioneering work in pulsed optical and infrared SETI including the NIROSETI project and the Panoramic SETI (PANOSETI) project. At the ceremony the SETI Institute also honored the recipients of the 2020 and 2021 SETI Forward Awards, given for undergradauate work in SETI and astrobiology.  The recipients were Karen Perez, Ellie White, and Siddhant Sharma.

In May, the community learned that Franklin Antonio had suddenly and unexpectedly passed.  Best known as the co-founder of the semiconductor company Qualcomm, Antonio was a supporter of SETI Institute, and was a funder and participant in the PANOSETI project at UC San Diego. In 2023 it would be revealed that Antonio left a \$200 million gift to the SETI Institute in his will, a transformative gift that would ensure the Allen Telescope Array and research efforts at the SETI Institute can continue and grow.

In September, the community received sad news that Frank Drake, arguably the founder of SETI, passed away at age 92. Drake's contributions are hard to overstate, and include the first SETI search \citep{OZMA}, organizing the scientific program for the 1961 Green Bank Conference at which he wrote down the Drake Equation \citep{DrakeEquation}, the construction of the Arecibo Message \citep{AreciboMessage}, contributions to the Pioneer Plaques and Voyager Golden Records \citep{SaganMurmurs}, and numerous contributions to the problem of frequency selection and survey design. He also served as the director of the Carl Sagan Center and as chairman of the Board of Trustees for the SETI Institute. 
Numerous obituaries outlined his extraordinary career, including \citet{Scoles2022}, \citet{Frank2023}, and \citet{Billings_obit}. 

And in October the community learned that early career SETI researcher Bart Wlodarczyk-Sroka died unexpectedly while visiting and collaborating with the Breakthrough Listen team in California.  Wlodarczyk-Sroka earned his masters degree with Michael Garrett at the University of Manchester and was a rising star in the field.

\section{Results from Searches (\thesearches\ papers)}

\subsection{Breakthrough Listen}
Continuing its leadership in the field, the Breakthrough Listen project provided new search progress in four papers. \citet{Garrett2022} explored the over 140,000 background galaxies and other objects in the beams of Breakthrough Listen radio observations at Green Bank Observatory.  They pointed out that the already published null results also rule out powerful transmitters from these sources, and so provide a very tight upper limit on the rate of transmitters with EIRP\footnote{EIRP is Equivalent Isotropic Radiated Power and is a measure of how strong a signal is, equal to the actual radiated power times the gain of the transmitting dish.  The EIRP of the Arecibo interplanetary radar was $6\times 10^14$ W} values of order $10^{25}$ W.
 
\citet{Gajjar2022}, noting the efficiency of pulsed signals compared to continuous wave narrowband signals, deployed a GPU-driven convolutional neural network to search archival high-time-resolution spectra taken from  4--8 GHz of 1,884 stars. \citet{Franz2022} presented null results for observations taken during transits for 61 TESS transiting planets from 1--11 GHz, with an upper limit EIRP of around $10^{14}$ W.
\citet{Perez2022a} described a radio study of 2MASS 19281982-2640123, the candidate star identified by \citet{Caballero2022a} (TF) using Gaia as a nearby Sun-like star in the Wow! Signal footprint.

\subsection{Searches at the Solar Gravitational Lens}

Collaborating with Breakthrough Listen,~\citetalias{Tusay2022}~\citep{Tusay2022} explained how one might search for signals sent towards the inner solar system from an interstellar communication network of relay probes at the
solar gravitational foci of nearby stars. They report a null results for a search targetting probes at the foci for {{$\alpha$}} Centauri in L and S bands.  

Following a similar line of reasoning,~\citet{Gillon2022} reported a null detection of optical signals from a hypothetical transmitter sending signals through the solar gravitational lens towards Wolf 359 during the time
the Earth was in a position to have intercepted such transmissions. 

\subsection{VASCO}

The Vanishing \& Appearing Sources during a Century of Observations (VASCO) project reported three results in 2022 based on archival photographic plates. \citet{Villarroel2022c} (IM) described a search strategy to identify objects orbiting Earth in geosynchronous orbits in pre-1957 images from their glints. \citet{Villarroel2022d} reported the results of the search, finding some ambiguous but intriguing candidates, and reporting the tight upper limit that would result once their inconclusive candidates could be ruled out.

\citet{Solano2022} reported a null result for vanishing sources in archival POSS images. As an ancillary result, they report a low rate of failed supernovae in the Milky Way.

\subsection{Other Searches}

In the optical, \citet{Suazo2022} combined optical photometry and parallaxes from Gaia DR2 with mid-infrared data from ALLWISE to set tight upper limits on Dyson spheres in the Milky Way.
	
Some of the first results from the FAST telescope in China were reported by \citet{Tao2022}, who performed 1.05--1.45 GHz observations toward 33 known exoplanet systems, with EIRP upper limits of $\sim10^9$ W. Using the Murchison Widefield Array, \citet{Tremblay2022} reported a null result at 155 MHz towards the Galactic Center and bulge, having searched a huge number of stars at EIRP upper limits of $\sim10^{19}$ W. 

\citet{Schmidt2022} extended a previous search for analogues to Boyajian's Star to much of the northern sky, finding 15 additional candidates, and suggesting that their distribution across the HR diagram warrants SETI interest.

\citet{Zuckerman2022b} used the results of prior searches for infrared execesses and optical variability of white dwarfs to put upper limits on Dyson spheres around white dwarfs.

\citet{Kipping2022} explored the possible repetition timescales for the Wow! Signal under the assumption that it is a stochastically repeating signal, recommending 60 days of additional monitoring to rule out the possibility.

\subsection{\oumuamua}

 Conversations around the interstellar object \oumuamua continued in 2022. \citet{Loeb2022} discussed the possibility of ʻOumuama’s extraterrestrial origin, arguing it could be a light craft pushed by sunlight. \citet{Zhou2022} calculated the orbital and photometric signatures of a thin, tumbling craft and concluded they were inconsistent with \oumuamua's observed properties.

 \citet{Matarese2022} (T) considered a meta-empirical approach to Loeb's hypothesis and discussed the nature of the question. \citet{Lineweaver2022} (T) argued the likelihood \oumuamua was an alien craft depends strongly on one's priors, and that the appropriate priors are that extraterrestrial life is unlikely to exist.  
 \citet{Zuckerman2022a} (T) argued that the motivations for sending such a probe are so unlikely that it must have been a natural object.  
 
 \citet{Elvis2022} (DT) discussed possible motivations for sending the probe to Earth, and research programs they might motivate for future interstellar objects.

\section{Search Methods and Instrumentation (\themethods\ papers)}

\subsection{Instrumentation}

An exciting upcoming project is the next generation of the Very Large Array (ngVLA), a powerful addition to radio SETI. The optimal antenna array and its implications for SETI were described by \citet{Ng2022}.

\citet{Maire2022} gave an update on the PANOSETI project, including its first detection of astrophysical gamma rays via Cherenkov radiation, and calibration in collaboration with the VERITAS experiment.

\citet{Osmanov2022} explore several prospects for using small-sized optical telescopes for SETI, including the possibility of detecting hot Dyson spheres and Type-3 civilizations.

\citet{Haqq-Misra2022_Technoclimes} synthesized the results of the 2020 Technoclimes workshop, focusing on the the detectability of a wide range of non-radio technosignatures with current and future facilities.  \citet{Haqq-Misra2022_Astro2020} and \citet{Haqq-Misra2022_PS_Decadal} similarly identified language in the decadal surveys of astrophysics and planetary science that justifies technosignature searches, and opportunities in the recommended hardware programs of those reports for SETI.
 
\subsection{Computation}

\citet{Pinchuk2022} presented a novel convolutional neural network approach for radio frequency interference mitigation in the search for radio technosignatures, reducing the need for visual inspection of candidates in dynamic spectra by a factor of 6--16. \citet{Chen2022b} outline a method of enhancing signal detection in radio dynamic spectra with edge detection methods.

\citet{Brzycki2022} described \texttt{setigen}, a Python-based library for injecting artificial signals into radio dynamic spectra for injection-recovery and other tests of SETI search algorithms.

\citet{Socas-Navarro_Berea} presented an overview of searches for megastructures in time-series data and the various methods for finding them.

\section{Target and Frequency Selection (\thetargets\ papers)}

\citet{Davis2022} proposed that an analysis of exoplanet host star positions could allow us to deduce the topology of an interstellar communication network, and focus efforts on searching for broadband microwave emission coincident with X-ray timing information from likely hubs in the network.

\citet{Suphapolthaworn2022} calculated the observability of Earth from other stars via gravitational microlensing, and finds it to be quite low with Earth-level technology. They define the Earth Microlensing Zone and recommend SETI searches where it overlaps the Earth Transit Zone near the Galactic center.

\citet{Davenport2022} detailed the search methodology of using the SETI Ellipsoid to find signals coordinated temporally with salient events, and applied it to the case of SN 1987A.


\section{Development of Technosignatures (\thedevelopment\ papers)}

\subsection{Communication Forms}

Building on the pioneering work of \citet{freudenthal1960lincos}, \citet{Matessa2022} detailed a method of decoding of interstellar ETI messages by looking for the structure of a message, locating and defining characters and symbols, patterns, ratios, and expressions of physics and mathematics. \citet{McConnell_Berea} provided an overview of how message decoding and interpretation might proceed, including some of the material from his 2021 book \citep{mcconnell2021alien}.

In the first in a series of papers, \citet{Crilly2022a,Crilly2022b} hypothesized that alien signals will exhibit delta-t delta-f repetition to distinguish themselves from background noise, and described how communication signals showing quantized patterns would be detectable by radio telescopes.

\citet{Li2022} explored how the orbit and rotation of an exoplanet would contribute to the drift rate of radio signals transmitted from them. Exploring the optics of the solar gravitational lens, \citet{Engeli2022} showed that even a 1W laser pointer at Proxima could be detectable from the solar system through the lens.

\subsection{SETI and Stellar Evolution}

\citet{Huston2022c} studied the evolutionary and observational consequences of a Dyson sphere's radiative feedback on a star, showing they would be important for very hot or very complete Dyson spheres.  \citet{Scoggins2022} detailed a process in which an advanced civilization might 
siphon mass from its host star as it ages to counteract its natural increase in luminosity, prolonging the habitable lifetime of its planetary system.

\subsection{Atmospheric and Geological Technosignatures}
 \citet{Beatty2022} calculated the detectability of city lights upon the night sides of habitable worlds for upcoming space based direct imaging missions such as LUVOIR (which has implications for Habitable Worlds Observatory). \citet{Haqq-Misra2022_Nitrogen} suggested using the simultaneous presence of \ce{NH3} and \ce{N2O} alongside \ce{H2O}, \ce{O2}, and \ce{CO2} in exoplanetary atmospheres could be a sign of industrial farming and production of fertilizer.

\citet{Haqq-Misra2022_CFCs} calculated the detectability of chlorofluorocarbons in exoplanetary atmospheres, concluding that under very optimistic assumptions, JWST could detect high levels of CFCs in the atmospheres of habitable zone M dwarf exoplanets like TRAPPIST-1\textit{e}.

\subsection{Other Novel Technosignatures}
\citet{Embaid2022} discussed the presence of heideite and brezinaite in meteorites and noted their similarities to certain superconductors, raising the question of whether they might be technosignatures.

\citet{Berera2022} discussed the viability of quantum communication over vast interstellar distances, and explored how to detect and possibly interpret such signals.

\citet{Clement2022} described exoplanetary monumental architecture as a technosignature in the form of arranging exoplanetary systems into series of unnatural, perfect mean-motion resonances.

\citet{Vukotic_Berea} explored the possibility of habitable ``mini-Earths,'' or shells of material $\sim$100 km in radius around a $\sim 10^{20}$ kg ($10^{-5} M_\oplus$) black hole used to generate $\sim 1$g of surface gravity.

\section{Theory of ETIs (\thetheory\ papers)}

\subsection{The Drake Equation \& the Fermi Paradox}

\citet{Wright2022c} used a pair of Drake-like equations for biosignatures and technosignatures to argue that technosignatures might be more abundant, longer-lived, more detectable, and less ambiguous than biosignatures, and argued that both approaches should be pursued.  \citet{Impey2022} came to the opposite conclusion: that technosignatures are the least likely to succeed, but also noted that searches for them are the most challenging to assess.

\citet{Dobler2022} discussed the solutions to the Fermi Paradox and their implications for technosignature searches, noting that optimistic explanations for the lack of contact to date often invoke assumptions that are pessimistic with respect to the prospects of future contact.  \citet{Lindsey2022} used an intercivilizational political lens to examine how assumptions of a ``unified rational actor,'' already critiqued in political science, are likely equally unwarranted in hypotheses about the behavior of ETIs, and the resulting implications for the Fermi Paradox.

\citet{Wandel2022} suggested a solution to the Fermi Paradox that ETIs universally wait for the “Contact Era”---i.e.\ the advent of radio technosignatures on a planet---before initiating physical contact.  

\citet{Schleicher2022} described the rise and fall of life and civilizations in the Galaxy in a set of coupled differential equations, and use them to explore the implications of various factors such as civilization lifetime and rate of abiogenesis, concluding the distance to the nearest ETI is highly uncertain.  \citet{Song2022} simulated the history of the Galaxy under different assumptions of the time and chance needed for communicative species to evolve, and calculated the timescales for one- and two-way communication to be established among these species. \citet{Burns2022} performed a calculation focusing on existential threats, concluding that the astrophysical extinction of civilizations happens roughly once every 100 million years, and that the most common cause of extinction is meteorite impact.

\citet{daSilva2022} reviewed various Drake-like Equations for the abundance of life in the universe and developed their own formalism for classifying the stages of a species' development towards galactic settlement, arguing that there exists a bottleneck near our level of technology that explains the Fermi Paradox and Great Silence.

\citet{Balbi_Berea} discussed the statistical issues surrounding SETI and the Drake Equation, and how to use Bayesian statistics to interpret upper limits from radio surveys.

\subsection{Interstellar Migration}

\citet{Gould2022} proposed a method for intergalactic travel via gravitational slingshots using stellar mass black holes orbiting supermassive black holes, as well as intragalactic travel via tight white dwarf binaries. Gould argued these methods of travel strengthen the Fermi Paradox.

\citet{Romanovskaya2022b} discussed how advanced extraterrestrials could utilize free-floating planets to explore other star-systems, and \citet{Romanovskaya2022a} (DT) proposed possible technosignatures these ``Cosmic Hitchhikers'' might produce along their journey, as well as strategies to search for these artifacts. 

\citet{Haqq-Misra2022_LowMass} offered a solution to the Fermi paradox invoking a preference for all expanding civilizations to favor low-mass stars over stars like the Earth.

\citet{Davis_Berea} suggested determining the optimal form of interstellar communication networks from network and graph theory considerations of the spatial distribution of known exoplanets.

\subsection{Interstellar Probes}

\citet{Ellery2022b} discussed the role of self-replicating probes in the Fermi Paradox, arguing that it would be anti-Copernican to think other species would not build them, since humanity is close to building them.  Ellery argues certain types of clay in the solar system would be evidence of visitation by such probes.

\citet{Matloff2022} gave several reasons a civilization would build self-replicating probes, discussing how they would propel themselves, what their launch strategies would be, and where the Solar System to search form them.

\citet{Chen2022a} looked at a common objections to self-replicating probes, that they would mutate and evolve away from exponential growth, arguing that a highly successful mutation would replace prior species and continue exponential growth.

\citet{Ezell2022} modeled the density of spacecraft in the Galaxy and developed a formalism for inferring the launch rates of interstellar objects by ETIs based on the detection of interstellar objects in the Solar System.

\citet{Smith_2023} discussed the trade-off between the sophistication of interstellar probes and their launch dates, arguing that the first to arrive at a system might be the most advanced.

\subsection{Other theory}

\citet{Caballero2022b} gave an estimate for the prevalence of hostile ETIs in the Galaxy based on our own history of warfare, and \citet{Taylor2022} offered a rebuttal criticizing the methodology.

\citet{Frank2022} suggested intelligence is not only an individual property, but is best understood and observable as a planetary-scale transition whose trajectory can be modeled.

\citet{Berea_Berea} provided a thorough investigation of the many ways in which work from the field of economics can inform various aspects of SETI.

\section{Social aspects of SETI (\thesocial\ papers)}

\citet{Charbonneau_Berea} provided a historical analysis of the origins of SETI in the US and USSR through the lens of the Cold War, including the origins of the astrobiology and SETI classic \textit{Intelligent Life in the Universe} by \citet{shklovskii1966}.

\citet{Wright2022a} critiqued the ``realpolitik'' analysis of \citet{Wisian2020} of the terrestrial risks of a successful SETI detection in terms of geopolitical destabilization.  They find these concerns to be based on a narrow and contrived contact scenario, and advocated for transparency and education of decision makers as antidotes to the concern. \citet{Davis_Shillo_Berea} discussed issues surrounding the cultural impact of a successful detection, including the diffusion of the news into society.

\citet{Szocik2022} discussed why ETIs might be unable to recognize humans as fellow intelligent beings, and how a feminist lens challenges suggestions that humans might share a common basis in mathematics or engineering that would aid message comprehension.

\citet{DeVito_Berea} discussed the nature of universal message composition, and what the nature of science can teach us about extraterrestrial communication and science.  \citet{Jiang2022} presented a binary-coded message as an update to the Arecibo Message and suggest times for its transmission, ostensibly by FAST or the Allen Telescope Array.

\section{Missed in 2021}

In an effort to be more comprehensive in our catalogue, we reviewed our process and noticed that 9 relevant SETI publications were missed in 2021 by \citet{Huston2022a}, primarily published in the Journal of the British Interplanetary Society (JBIS). A small addendum to SETI in 2021, we sort them here:

\subsection{Results from Searches}

\citet{Benford:2021:_Wow!} suggested that the ``Wow! Signal" may have been unintentional leakage from a distant artificial beam, such as a pulsed laser for solar sail propulsion, and all-sky surveys in microwave and laser bands may discover more such transient signals.

\citet{Sheerin:2021:} considered whether a solar thermal propulsion vehicle could fit as an alternative explanation for the interstellar object \oumuamua.

\subsection{Development of Technosignatures}

\citet{Baxter:2021:} considered the detectability of a speculative dark energy ram jet engine that could be used for intergalactic travel or to mitigate localized cosmic expansion around galaxy clusters.

\citet{Kezerashvili:2021:} highlighted the ability of the \textit{Gaia} mission to confirm or refute a previous suggestion that older stars in the Milky Way seem to be suffering anomalous accelerations, and mention stellar engines as a possible explanation.

\subsection{Theory}

\citet{Ashworth:2021:} argued that the evolution of intelligent activity would necessarily lead to galactic colonization and the current lack of evidence for such a scenario implies that process is either in its earliest phases or has not yet begun.

\citet{Benford:2021:_Probes} estimated the number of close stellar encounters in our Solar System's history to motivate the Search for Extraterrestrial Artifacts (SETA) on or around the Moon and Earth Trojans. Benford argued that potential civilizations around the stars that get within about a light-year may be able to discern an ecosystem on Earth and establish long lived probes to monitor our planet.

\citet{Fow:2021:} examined the failure modes of self-replicating probes to show that implementing such technology might not necessarily lead to the eventual galaxy-wide population that is often seen as inevitable and commonly linked to the Fermi Paradox.

\citet{Robertshaw:2021:} linked the cyclical crossing of the galactic mid-plane with cyclical global extinction events, suggesting that these events are necessary for the eventual emergence of intelligence in a biosphere. They considered the parameters that place a star within this galactic habitable zone and suggested a search strategy to compare signals from these galactic regions to others as indirect evidence of the emergence of technological species.

As part of a collection on multimessenger astrophysics, \citet{Vukotic:2021:} surveyed modern SETI methodologies across and beyond the EM spectrum.

